\documentclass[conference]{IEEEtran}
\usepackage[hyphens]{url}
\usepackage{hyperref}
\hypersetup{breaklinks=true, colorlinks,allcolors=blue}
\usepackage[backend=bibtex,style=ieee,citestyle=ieee-comp]{biblatex}

\IEEEoverridecommandlockouts
\usepackage{amsmath,amssymb,amsfonts}
\usepackage{algorithmic}
\usepackage{graphicx}
\usepackage{textcomp}
\usepackage{xcolor}  % for colored inline text
\usepackage{balance}
\usepackage{mathtools}  % for cartesian product symbolb bigtimes
\usepackage{orcidlink}  % for orcid in author list if wanted
\usepackage{kantlipsum} % used to generate filler text
\usepackage{csquotes} %use for correct quotation mark directions
\usepackage{makecell}
\setcellgapes{2pt} % ensure spacing around the makecells
\makegapedcells
\usepackage{array} % for mid alignment in tables
\usepackage{soul} % for strikethrough, can be removed later
\usepackage{comment} % for block comments, can be removed later
\usepackage{enumitem} %for custom enumerate numbering
\usepackage{balance} % for balancing columns on last page

\def\BibTeX{{\rm B\kern-.05em{\sc i\kern-.025em b}\kern-.08em
    T\kern-.1667em\lower.7ex\hbox{E}\kern-.125emX}}

\begin{document}

\title{Assessor Experiences in CMMC Level 2 Certification Assessments: An Interpretative Phenomenological Analysis of Role Expectations}

\author{\IEEEauthorblockN{Samuel Heuchert\,\orcidlink{0000-0001-5086-6046}}
\IEEEauthorblockA{\textit{The Beacom College of Computer and Cyber Sciences} \\
\textit{Dakota State University}\\
Madison, SD, USA \\
samuel.heuchert@trojans.dsu.edu}
\and
\IEEEauthorblockN{John Hastings\,\orcidlink{0000-0003-0871-3622}}
\IEEEauthorblockA{\textit{The Beacom College of Computer and Cyber Sciences} \\
\textit{Dakota State University}\\
Madison, SD, USA \\
john.hastings@dsu.edu}
}

\maketitle

\begin{abstract}
The Cybersecurity Maturity Model Certification program requires third-party assessments be conducted under a non-consultative model. The model is intended to ensure impartiality for organizations seeking certification. While this structure defines expectations for assessor behavior, assessor experiences and interpretations of these constraints remain underexamined. The study examines the lived experiences of CMMC-Certified Assessors and how they navigate role expectations within the non-consultative model. Using Role Conflict Theory as a guiding framework, Interpretative Phenomenological Analysis (IPA) was applied to semi-structured interviews to explore how assessors make sense of their roles. The analysis identified experiential themes that describe how assessors construct professional credibility, execute structured assessment work, and manage the practical challenges of maintaining non-consultative boundaries. Findings indicate that assessors rely on technical competence, procedural discipline, and boundary management strategies to reconcile competing expectations. As an exploratory study, the results are not intended to be generalizable but provide initial empirical insight into assessor experiences, highlight considerations related to boundary clarity and assessor/organization interaction, and demonstrate the suitability of
IPA for examining practitioner experience within cybersecurity compliance contexts.
\end{abstract}


\section{Introduction}
The Cybersecurity Maturity Model Certification (CMMC) program is designed to verify that Department of Defense (DoD) contractors protect Federal Contract Information (FCI) and Controlled Unclassified Information (CUI). At 
Level 2, CMMC requires an organization seeking certification (OSC) to undergo an assessment by a third-party organization known as CMMC Third‑Party Assessment Organization (C3PAO) \cite{office_of_the_secretary_of_defense_32_2024}. Assessors play a central role in determining compliance outcomes in the CMMC program.  Their evaluations determine whether OSCs meet the required security controls. 

The assessors employed by C3PAOs are prohibited from providing consultative support to the OSCs in order to maintain impartiality \cite{cyberab_cmmc_2024} while evaluating security requirements aligned with the National Institute of Standards and Technology (NIST) Special Publication (SP) 800-171 Rev. 2 \cite{ross_protecting_2021, ross_assessing_2018}. This  consulting prohibition may create role tension where assessors must remain neutral while also responding to practical assessment needs in complex environments. 

Although CMMC policy defines expectations for assessor impartiality, less is known about how assessors experience, interpret, and enact these expectations during assessment practice. The perspective of assessors operating within this non-consultative model has not been examined in existing CMMC research. Understanding assessor experiences clarifies how policy requirements are operationalized during assessments and how constraints, such as the non-consultative model, shape assessment practices. 

This study examines assessors' lived experiences as they interpret and fulfill role expectations within the non-consultative CMMC assessment model. It applies a qualitative approach to examine how assessors interpret role expectations, role constraints, and assessment practices.
The study's findings provide early insight into how assessor role expectations are interpreted and enacted in practice, clarifying how policy constraints may shape assessor behavior. 
The study is guided by the following primary research question:       

\begin{enumerate}[label={\textbf{RQ:}},left=1.0em]
            \item How do CMMC assessors experience and interpret role expectations during Level 2 Certification Assessments?
            \end{enumerate}

The remainder of this paper is organized as follows. Section \ref{background} provides background on CMMC and the non-consultative assessment model. Section \ref{litreview} reviews related literature. Section \ref{methodology} describes the methodology. Section \ref{results} presents the findings. Section \ref{discussion} discusses implications, contributions, and limitations. Section \ref{conclusion} concludes the paper.

\section{Background}
\label{background}

    \subsection{CMMC Policy Intent}
        The DoD created the CMMC program to improve the security posture of the Defense Industrial Base (DIB) and to ensure that FCI and CUI are protected within the defense supply chain \cite{united_states_congress_national_2019}. 
        The program requires contractors to demonstrate implementation of specified security controls \cite{office_of_the_secretary_of_defense_32_2024, u_s_department_of_defense_dfars_2017} and was designed to promote consistent and repeatable assessment outcomes across the defense industrial base. 
        
        At Level 2, the program includes two assessment paths: self-assessments and third-party assessments. A successful self-assessment results in a Level 2 (Self) status for the Organization Seeking Assessment (OSA). A third-party assessment conducted by an authorized C3PAO results in a Level 2 (C3PAO) status for the OSC, which is a subset of an OSA. Assessors employed by C3PAOs in the third-party path play a central role in verifying control implementation and determining compliance.
        
        The requirement to implement the security controls has been in effect since 2016. The CMMC Program is validating the implementation of the requirements because previous self-attested models did not provide adequate assurance to the DoD \cite{u_s_department_of_defense_dfars_2017, office_of_the_secretary_of_defense_32_2024, sera-brynn_reality_2019, sera-brynn_reality_2020}. 

\subsection{Assessor Impartiality \& the Non-Consultative Model}
    %\subsection{Policy Design}
        The CMMC Program Rule, 32 CFR 170 \cite{office_of_the_secretary_of_defense_32_2024} requires a Code of Professional Conduct (CoPC) \cite{cyberab_cmmc_2024} to establish the rules that govern assessor impartiality. The CoPC prohibits assessors from providing consultative support to OSCs in order to prevent conflicts of interest and maintain impartiality as part of the program's broader design to maintain assessment independence and integrity. 
        
        Impartiality is maintained through explicit limits on the type and extent of engagement assessors may have with OSCs. The non-consultative model prevents assessors from providing remediation guidance to OSCs, which may leave organizations without clear direction for corrective action on deficient security controls. 
        %new
        These limits define the practical boundary conditions under which assessors must perform assessment work.

\subsection{Practical Role Tensions in Assessment Work}
The CMMC non-consultative assessment model reflects a common regulatory challenge of translating policy intent into consistent operational practice \cite{chriqui_advancing_2023}. In this context, assessors must apply impartiality requirements while interacting with OSCs that may seek clarification, feedback, or guidance during assessment activity. These tensions are treated in this study as part of the assessment context rather than as a full theoretical explanation; relevant role-based theories are reviewed in Section \ref{role_conflict}.

\section{Literature Review}
\label{litreview}
    \subsection{Audit and Assurance Foundations}
    Audit and assurance systems are designed to provide independent verification of organizational practices and to establish confidence in compliance outcomes across regulatory environments \cite{turley_corporate_2004, srivastava_bayesian_2009, manning_triangulation_2018}.  Established frameworks emphasize the importance of auditor independence, evidence evaluation, and methodological rigor in producing reliable and consistent assessments \cite{stewart_internal_2010, srivastava_bayesian_2009, noauthor_au-c_2023}. Standards such as ISA 530 and ISO/IEC 27007 \cite{international_auditing_and_assurance_standards_board_iaasb_international_2009, international_organization_for_standardization_iso_isoiec_2020, noauthor_au-c_2023} highlight the role of risk-based sampling, documentation sufficiency, and professional judgment in determining whether sufficient evidence exists to support assurance conclusions \cite{danescu_professional_2014, kamau_factors_2018}. These principles establish the foundational role of assessors in translating formal requirements into operational verification processes \cite{coetzee_use_2013, manning_triangulation_2018, stewart_internal_2010}.
    
    \subsection{Independence, judgment, and Variability in Assurance Systems}
    A recurring challenge in assurance systems involves balancing independence with the desire to support organizational improvement \cite{gramling_consulting_2018, mahieux_auditors_2022, ferrari_management_2023}. Power's \cite{power_audit_1997} analysis of audit independence shows that assurance mechanisms often struggle to separate verification from advisory functions.
     
    Empirical work further demonstrates that assessor decision-making is not fully standardized \cite{laghmouch_auditors_2024, detzen_impact_2024, hall_auditor_2023}. Therrien and Hastings \cite{therrien_need_2026} found that evidence sampling in CMMC assessments is frequently driven by assessor judgment, perceived risk, and environmental complexity rather than standardized methodologies.
     
    Together, these findings indicate that independence constraints, professional judgment, and methodological variability are structurally embedded features of assurance systems \cite{cameran_audit_2018, honkamaki_homogeneity_nodate, ahn_effect_2017}. However, existing research does not examine how individuals operating in assessor roles interpret and navigate these conditions in practice. While these challenges are well documented in traditional audit and assurance domains, they remain underexamined in cybersecurity-specific compliance environments \cite{rindasu_information_2016, kizirian_influence_2004, coles-kemp_information_2006}.
  
    \subsection{Cybersecurity Compliance Frameworks}
    Research on cybersecurity compliance frameworks primarily examines organizational implementation, regulatory intent, and oversight structures rather than the experiences of assessors \cite{moeti_compliance_2025}. Studies of compliance verification programs focus on how organizations adopt controls, manage documentation, and respond to certification requirements \cite{folorunso_security_2024, moeti_compliance_2025}. Prior research on compliance regimes such as International Standards Organization (ISO) 27001, Health Insurance Portability and Accountability Act (HIPAA), and System and Organization Controls (SOC) focuses primarily on organizational experiences and outcomes rather than on assessors' experiences \cite{folorunso_security_2024}.

    Prior work identifies a persistent gap between policy intent and operational outcomes, noting that compliance structures often fail to translate regulatory goals into consistent practice \cite{solove_information_2018, moeti_compliance_2025, folorunso_security_2024}. This gap is reflected in findings that organizations frequently prioritize validation over substantive security improvement, resulting in compliance being treated as a checkbox activity \cite{folorunso_security_2024}. This limitation is particularly significant in environments where assessors operate under strict independence constraints, such as the non-consultative model used in CMMC assessments, where interpretation must occur without the ability to provide guidance.

    \subsection{Evolution of CMMC}
    CMMC emerged from broader federal efforts to strengthen information security and supply chain assurance. The Federal Information Security Management Act established post-9/11 requirements for federal agencies to implement standardized information security controls \cite{united_states_congress_44_2009}. Subsequent supply chain incidents revealed significant variability in contractor compliance, prompting the Department of Defense (DoD) to introduce CMMC as a shift from self-attestation to independent verification \cite{u_s_department_of_defense_dfars_2017, hoesman_giving_2023, sera-brynn_reality_2020}.

    Existing research on CMMC focuses on contractor readiness, cost implications, implementation challenges, common deficiencies \cite{sundararajan_most_2022} and variability in assessment practices \cite{therrien_need_2026}. While the work provides insight into how the framework operates at an organizational and process level, it does not address the experiences of assessors who perform verification. This shift toward independent verification places assessors at the center of compliance assurance, thereby formalizing their responsibility for determining certification outcomes while leaving their role interpretation underexamined.
    
    \subsection{Role Conflict Theory in Organizational Contexts}
    \label{role_conflict}
    Role conflict occurs when individuals are subject to incompatible or competing expectations within a defined role, making Role Conflict Theory \cite{kahn_conflict_1964} a useful lens for examining how assessors interpret role expectations in constrained environments. Adding to Role Conflict Theory is Organizational Role Theory, which explains how these role expectations shape behavior within institutional systems and influence role performance \cite{katz_social_1978}. Core constructs within this framework include role ambiguity, role conflict, and role negotiation, which describe how individuals interpret, reconcile, and respond to competing expectations in structured environments.
    
    Prior applications of the theories focus on clinical, educational, and managerial environments, where competing demands or conflicting expectations influence professional performance. Role Conflict Theory has not been applied to cybersecurity compliance or assessment roles, particularly in environments with formal independence constraints. Studying assessor experiences within CMMC provides an opportunity to extend role conflict theory into a new domain where regulatory constraints, impartiality requirements, and operational demands interact in ways not examined in existing literature.
    
    \subsection{Synthesis and Research Gap}
    Across the literature, audit and assurance research establishes that verification systems are shaped by independence requirements, professional judgment, and methodological variability. However, cybersecurity compliance and CMMC research primarily examine organizational implementation, framework design, readiness, and assessment variability. What remains underexamined is how assessors experience and interpret role expectations under formal independence constraints. This omission is significant in CMMC because assessor interpretation directly influences how the non-consultative model is enacted during Level 2 Certification Assessments.

\section{Methodology}
\label{methodology}

The study used an exploratory qualitative design to examine how assessors experience and interpret role expectations within the non-consultative model of CMMC Level 2 assessments. The design prioritized depth of understanding within individual accounts rather than generalization to the broader assessor population.
    
    \subsection{Participants and Sampling}
        Participants in the study were CMMC-Certified Assessors (CCA) with the Lead Assessor designation. Inclusion criteria required that each participant have participated in at least five assessments conducted under the non-consultative model and be qualified to serve as a Lead Assessor capable of directing assessment teams. This study included two participants.  Consistent with IPA, the small sample supports idiographic depth rather than generalization.
    
    \subsection{Data Collection}
        Data collection used semi-structured interviews conducted via Zoom. Participants were recruited through an 
        online recruitment post distributed via a professional Discord community of CMMC assessors. Each interview lasted approximately 60 minutes and was audio-recorded. Recordings were transcribed using Zoom's transcription functionality and reviewed for accuracy prior to analysis.         
        The interview protocol included open-ended prompts designed to elicit participant reflections on role expectations, operational constraints, and the interpretation of non-consultative requirements. The semi-structured format allowed participants to guide the discussion while ensuring coverage of core research topics.
    
    \subsection{Data Analysis}
    Interview transcripts were analyzed using Interpretative Phenomenological Analysis (IPA), a qualitative analytic approach suited to examining how individuals make meaning of lived experiences within specific contexts and providing an in-depth understanding of the meaning-making processes \cite{smith_interpretative_2009}. The method supported idiographic, case-focused analysis and was appropriate for small samples that required detailed examination of individual accounts.
            
    The analysis followed an iterative, multi-stage approach that included repeated reading of transcripts, initial noting, development of emergent themes, and identification of connections within each case. Coding and theme development were conducted manually using Microsoft Excel to support systematic organization of transcript excerpts, notes, and theme structures.

    Personal Experiential Themes (PETs) were developed for each participant based on close engagement with the transcript and iterative refinement of theme definitions. After individual case analysis, cross-case comparison was conducted to identify patterns of convergence and divergence, resulting in the development of Group Experiential Themes (GETs). The analytic process emphasized idiographic depth prior to abstraction across cases, consistent with IPA methodology. After the PETs and GETs were developed, Role Conflict Theory and Organizational Role Theory were used as interpretive frameworks to examine how the identified themes related to role ambiguity, role conflict, role negotiation, and structural role expectations.
    
    \subsection{Ethical Considerations}
    Participants were cybersecurity professionals working in a defense-adjacent environment. The research received Institutional Review Board approval (DSUIRB-20251014-02) prior to participant recruitment.  Confidentiality was protected through informed consent and the de-identification of all interview transcripts. Data was stored securely with access limited to the researchers. Participation was voluntary, and participants could decline to answer questions or withdraw at any time.   Participants were assigned aliases and are presented as ``Timothy'' and ``John'' in the findings.

    \subsection{Trustworthiness and Analytic Rigor}
    The study maintained an audit trail documenting analytic decisions, theme development, and movement from transcript excerpts to PET and GET structures. Theme development was repeatedly reviewed against the original transcripts to ensure alignment with participant meaning-making. Analytic reflexivity was maintained by distinguishing participant accounts from interpretive claims, consistent with IPA practices.

\section{Results}
\label{results}
In answering \textbf{RQ}, findings are presented first by participant PETs, followed by cross-case GETs. For each participant, a brief contextual description is followed by a table summarizing the participant's PETs, theme descriptions, and representative quotes.

    \subsection{Participant 1 - Timothy}
        Timothy is a CCA with a Lead CCA designation. His professional background includes several years of experience performing NIST SP 800-171 assessments, including work with historically Black colleges and universities. He has been recruited by multiple organizations to establish and develop their internal CMMC programs. Timothy has conducted more than 80 %eighty 
        NIST 800-171 assessments and has worked in roles involving coordination with assessment teams that include quality assurance reviewers, CCAs, and lead assessors. His experience spans the early, months-long CMMC assessments to more recent, streamlined processes of weeks-long assessments.  He demonstrated familiarity with both operational and organizational challenges associated with the non-consultative model of CMMC assessments. Timothy's PETs are summarized in Table \ref{pet_timothy}.
        
        \begin{table*}[t]
            \centering
            \caption{Personal Experiential Themes for Participant 1 (Timothy)}
            \label{pet_timothy}
            \begin{tabular}{p{1.8cm} p{5.2cm} p{9.2cm}}
            \hline
            \textbf{PET Name/ID} &
            \textbf{PET Description} &
            \textbf{Representative Quotes} \\
            \hline
            Competence as Credibility (A.1) &
            Credibility is understood as rooted in demonstrated competence, accumulated experience, and external validation. &
            \parbox[t]{9.2cm}{
                \textbullet\ \enquote{I received the credential based on my experience.}\\
                \textbullet\ \enquote{I have been headhunted by several organizations.}\\
                \textbullet\ \enquote{I have been brought in to start CMMC programs from the ground up.}
                }
                \\
            
            Evolving and Optimizable Assessment Work (A.2) &
            Assessment work is understood as evolving over time and capable of being optimized through experience and structural refinement. &
                \textbullet\ \enquote{The Level 2 certification assessment is clunky, not intuitive, and the wording takes a lot of work.}
            %\end{enumerate}
             \\%[4pt]
            
            Rules–Reality Tension (A.3) &
            Tension between formal non-consultative model rules and the practical realities of assessment work. 
            &
            \parbox[t]{9.2cm}{
            %\begin{enumerate}
                \textbullet\ \enquote{You almost need someone to consult with you for months to get ready.}\\
                \textbullet\ \enquote{In at least half of my assessments, clients have asked for consultation.}\\
                \textbullet\ \enquote{If you allow consultation, it becomes pay-to-play.}\\
                \textbullet\ \enquote{In FedRAMP, assessors can suggest things. I would not want that in CMMC.}
                }
            %\end{enumerate}
                \\
            \hline
            \end{tabular}
        \end{table*}

    \subsection{Participant 2 - John}
        John is a CCA with a Lead CCA designation. His professional background includes several years of experience working in the Risk Management Framework (RMF) before transitioning to CMMC, where he built a CMMC Program from the ground up.  He focuses on understanding the structural requirements governing assessor conduct under the CoPC. His account sheds light on the professional expectations placed on assessors, the importance of avoiding perceived conflicts of interest, and the practical challenges of maintaining clear boundaries between assessment and consulting activities. John's PETs are shown in Table \ref{pet_timothy}.

    \begin{table*}[t]
    \renewcommand{\arraystretch}{1.25}
    \centering
    \caption{Personal Experiential Themes for Participant 2 (John)}
    \label{pet_john}
    \begin{tabular}{p{1.8cm} p{5.2cm} p{9.2cm}}
    \hline
    \textbf{PET Name/ID} &
    \textbf{PET Description} &
    \textbf{Representative Quotes} \\
    \hline
    
    Structured Assessment (B.1) &
    Assessment work is experienced as a structured process requiring planning, documentation review, coordinated scheduling, and evidence management. &
    %\begin{enumerate}
    \parbox[t]{9.2cm}{
        \textbullet\ \enquote{As a Lead CCA, my role is to assist the OSC in preparing and planning for the assessment week.}\\
        \textbullet\ \enquote{I try to be specific, but staying on the assessment side has not been difficult for me.}
        }
    %\end{enumerate}
      \\%[4pt]
    
    Ethical Boundary Discipline (B.2) &
    Ethical discipline is centered on maintaining clear boundaries, avoiding real or perceived conflicts of interest, and ensuring transparency in assessor conduct. &
    \parbox[t]{9.2cm}{
    %\begin{enumerate}
        \textbullet\ \enquote{The key expectations relate to maintaining professionalism and integrity.}\\
        \textbullet\ \enquote{Any perceived conflict of interest, whether I notice it or someone else could perceive it, is something we must avoid under the Code of Professional Conduct.}\\
        \textbullet\ \enquote{Consulting is asked for by the OSC during the planning phase.}
        }
    %\end{enumerate}
       \\%[4pt]
    
    Boundary Maintenance Under Pressure (B.3) &
    Confident boundary maintenance in the face of OSC pressure is supported by the belief that third-party independence is necessary to prevent bias and uphold assessment integrity. &
    \parbox[t]{9.2cm}{
    %\begin{enumerate}
        \textbullet\ \enquote{I think the non-consultative requirement is productive. No matter how objective someone is, if you grade your own paper, it will look better than when someone else grades it.}
        }
    %\end{enumerate}
     \\
    \hline
    \end{tabular}
    \end{table*}

    \subsection{Group Experiential Themes}
        The GETs were developed through a cross-case comparison of the individual PETs. The purpose of this stage was not to generate a group norm or an average account of assessor experience. Instead, the analysis identified patterns of convergence and divergence across the contributing cases to understand how participants made meaning of similar role expectations in distinct ways. The GETs highlight shared experiential features while preserving each participant's unique perspective. The resulting themes reflect higher-level meanings that emerged only through comparison of individual cases and represent the collective experiential structure of the sample.
        
        Cross-case comparison of the PETs resulted in three GETs, as shown in Table \ref{table_get}. The themes synthesize convergences and divergences in how assessors interpret their roles, navigate boundary requirements, and engage with the operational demands of the non-consultative model that is CMMC.
    
    \begin{table}[ht]
    \renewcommand{\arraystretch}{1.25}
    \centering
    \caption{Group Experiential Themes}
    \label{table_get}
    \begin{tabular}{p{2.1cm} p{3.8cm} p{1.3cm}}
    \hline
    \textbf{GET Name/ID} &
    \textbf{GET Description} &
    \textbf{Contributing PETs} \\
    \hline
    
    Professional Credibility and Assessor Identity (C.1) &
    Shared emphasis on competence, legitimacy, and the role of expertise in maintaining assessor authority. & 
    A.1; B.1; B.2; B.3  \\[10pt]
    
    Assessment Work as a Technical, Evolving, and Highly Structured Practice (C.2) &
    Shared description of the assessment as procedural, evidence-driven, and continually developing. &
    A.2; B.3 \\[10pt]
    
    Navigating the Non-Consultative Model Boundary in Practice (C.3) &
    Shared perception that maintaining independence is necessary, but it is pressured and complex. &
    A.3; B.2; B.3 \\[10pt]
    
    \hline
    \end{tabular}
    \end{table}
    
    \subsubsection{GET C.1}
    Assessor identity is grounded in professional credibility. Across cases, participants emphasized competence, role clarity, and the disciplined execution of assessment responsibilities. Both Timothy and John framed credibility as something that must be demonstrated through performance, maintained through adherence to standards, and reinforced by alignment with the expectations of the CMMC program ecosystem. Assessor identity is understood as a function of expertise, training, and the ability to perform assessment work consistently and with integrity while being continually reinforced through the correct execution of assessment responsibilities.
        
    Divergence occurred when Timothy emphasized external validation, describing credibility in terms of recognition, recruitment, and organizational trust. John emphasized ethical validation, grounding identity in independence, boundary clarity, and adherence to professional conduct requirements.

\subsubsection{GET C.2}
    Participants described assessment work as a structured, procedural, and technically detailed activity that develops through experience and continual refinement. Across cases, there is an orientation toward planning, coordination, documentation, and the disciplined management of evidence and timelines. Assessment is not a static compliance exercise but a process that evolves within the ecosystem and requires sustained methodological precision.
               
    Divergence occurred when Timothy emphasized the potential for optimization and increased efficiency, focusing on how assessment practices can be refined and streamlined as the ecosystem matures. John emphasized structural discipline and adherence to processes, emphasizing the value of proper planning, procedural fidelity, and consistent execution throughout the full assessment cycle.

\subsubsection{GET C.3}
Participants described maintaining the non-consultative model boundary while managing the practical realities of assessment work. Across cases, assessors routinely encounter situations in which organizational expectations, preparation gaps, or requests for guidance create tension between what the rules permit and what the assessment context demands. Navigating this boundary is a core aspect of assessor work and requires ongoing attention to role clarity, transparency, and adherence to independence requirements.
        
Both participants emphasized that maintaining the line between assessment and consulting is central to the CMMC program's integrity. Both described OSCs as frequently pushing against this boundary, either through requests for consultation or attempts to define scope through assessor input. Both accounts highlight that the system relies on assessor independence and that assessors feel tension between practical needs in the field and strict non-consultative requirements.
        
Divergence occurred when Timothy focused on systemic contradictions and the operational burden imposed by formal rules that conflict with assessment realities. His account showed the friction caused by unprepared OSCs, the risk of pay-to-play dynamics, and the need to reject practices seen in adjacent information assurance compliance regimes. John emphasized clarity, confidence, and disciplined communication as strategies for boundary maintenance, framing independence as a matter of ethical discipline rather than structural contradiction.

\section{Discussion and Contributions}
\label{discussion}

    \subsection{Theoretical Interpretation of Findings}
    Interpreted through Role Conflict Theory and Organizational Role Theory, the three GETs show how assessor experience is shaped by both competing expectations and the institutional structures that define assessment work. Role Conflict Theory helps explain the tensions assessors experience when expectations are incompatible, such as when OSCs seek guidance while the non-consultative model prohibits consultation. Organizational Role Theory helps explain how assessor behavior within the CMMC ecosystem is shaped by formal program rules, professional standards, assessment procedures.
    
    GET C.1 reflects mechanisms for managing \textit{role ambiguity}. The emphasis on competence, experience, and adherence to professional standards functions as a stabilizing mechanism that reduces uncertainty in role expectations. These findings indicate that participants resolve ambiguity by grounding role interpretation in internally defined criteria for credible performance.

    GET C.2 aligns with \textit{role overload} and \textit{structural role complexity}. The procedural density and coordination demands are described in the presence of multiple concurrent role expectations. Participants' reliance on structured workflows and disciplined processes reflects adaptive responses to sustained workload pressure and complexity.

    GET C.3 most directly reflects \textit{role conflict} and \textit{role negotiation}. The incompatibility between organizational expectations for guidance and formal prohibitions on consultation represents a clear instance of conflicting role demands. The boundary management strategies described in the findings demonstrate how participants actively negotiate these constraints through communication, scope control, and reinforcement of independence requirements.

    Together, the findings show that assessors do not simply follow policy requirements mechanically. They interpret and enact their roles within a structured institutional environment while managing competing expectations in real time. This demonstrates how formal CMMC policy constraints translate into lived role tensions and adaptive professional behavior.

\subsection{Practical Implications}
Given the exploratory nature of the study, the following implications should be considered provisional rather than definitive program recommendations. The findings suggest three practical areas where CMMC program stakeholders, C3PAOs, assessors, and OSCs may reduce ambiguity and support more consistent assessment practice.

First, boundary expectations should be clarified before assessment activity begins. Participants described recurring situations in which OSCs sought clarification, feedback, or guidance that assessors could not provide under the non-consultative model.  This suggests that DoD CMMC program stakeholders and C3PAOs may benefit from clearer OSC-facing materials that explain what assessors can and cannot do during assessment interactions.  Formal pre-assessment orientation materials, role-boundary guidance, or standardized expectation-setting language may reduce misunderstandings that currently create boundary pressure during assessment work.

 Second, assessment preparation and evidence workflows should be further standardized.  Participants described Level 2 Certification Assessments as procedurally dense, evidence-driven, and dependent on coordinated documentation review, scheduling, and scope management.  Standardized scheduling templates, evidence submission guidance, domain-specific checklists, and structured pre-assessment review routines may reduce ambiguity for both assessors and OSCs.  These tools could also reduce variation in assessment execution that currently depends heavily on individual experience.

 Third, assessor training should continue to emphasize neutral communication and boundary-preserving assessment practice.  Participants described boundary maintenance as an ongoing practical task rather than a static rule.  Training focused on question framing, scope validation dialogue, and neutral evidence-gathering language may help assessors collect necessary information without creating the appearance of consultation.  Reinforcing professional expectations related to competence, ethical discipline, and CoPC-aligned conduct may also support consistency across C3PAOs and assessment teams.

 Together, these implications suggest that practical improvements do not require weakening the non-consultative model.  Instead, the findings point toward clearer role communication, more consistent assessment preparation, and stronger support for assessors as they navigate the boundary between evidence collection and consultation.

\subsection{Research Contributions}
This study makes three contributions to cybersecurity compliance research. First, it provides empirical insight into how CMMC assessors interpret role expectations, boundary constraints, and procedural demands within the non-consultative model. Second, the study contributes theoretically by illustrating how Role Conflict Theory and Organizational Role Theory can be applied to understand assessor experiences in a regulated assessment environment. Third, the study contributes methodologically by providing evidence supporting the use of IPA for examining practitioner experience in structured, policy-driven contexts, showing how idiographic analysis can capture role interpretation under formal constraints.

\subsection{Limitations \& Future Work}
The small sample size in this exploratory study was an inherent constraint of the design and limited the range of assessor perspectives represented in the analysis. Although idiographic depth is central to IPA, the findings reflect the experiences of two assessors and should not be generalized to the broader assessor population or to all organizational contexts within the CMMC ecosystem.
        
The study also reflects the perspectives of highly experienced assessors who demonstrate strong familiarity with assessment procedures and boundary requirements. Their accounts may differ from those of assessors with less exposure to complex assessments. The reliance on self-reported experience introduces the possibility that participants framed their role performance in ways that reflect professional norms or expectations rather than the full range of tensions encountered during practice.
        
The non-consultative requirement, assessment structure, and ecosystem dynamics explored in this study are unique to CMMC and may not align with assessor experiences in other cybersecurity or compliance frameworks. This context-specific structure shapes how role expectations and constraints are interpreted, potentially limiting transferability to adjacent cybersecurity compliance domains.

Future research should expand the participant sample to examine variation in assessor experience across organizational contexts and assessment environments. Additional focus on boundary pressure, organizational readiness, and the operational consequences of non-consultative requirements may provide further insight into how policy constraints influence assessment practice. Comparative analysis with adjacent compliance frameworks may also clarify how independence requirements shape assurance roles across domains.

\section{Conclusion}
\label{conclusion}
The CMMC program was established to ensure that defense contractors implement adequate protection for CUI. At Level 2 (C3PAO), this objective is operationalized through third-party assessments conducted by C3PAOs under a non-consultative model intended to preserve impartiality. Although the policy intent is well defined, limited research has examined how assessors interpret and enact their roles within these constraints. This study addresses this gap by documenting assessor meaning-making and identifying experiential patterns that reveal how independence requirements, operational demands, and role expectations interact during the assessment process.
    
    The research findings suggest that assessor experience is shaped by the interaction of role ambiguity, workload complexity, and competing expectations. Assessors rely on professional identity, structured practice, and boundary management strategies to maintain alignment with program requirements.
    
    The study contributes practically by providing early evidence relevant to policymakers, program designers, and C3PAOs. The findings highlight areas where the non-consultative model aligns with assurance objectives and areas where it creates operational tension that may influence the credibility and consistency of assessment outcomes. By centering on assessor experience, the study offers foundational insight into how CMMC program design affects those responsible for implementing its verification mechanisms and identifies domains where further refinement could support long-term program sustainability.

\section*{Acknowledgment}
A locally hosted large language model (LLM) was used for grammar and organizational assistance. All substantive content, analysis, interpretations, and conclusions are the authors' original work.

\balance
\printbibliography

\end{document}